\documentclass[prb,aps,twocolumn,preprintnumbers,amsmath,amssymb,superscriptaddress,showpacs]{revtex4-2}

\usepackage{graphicx}
\usepackage{epstopdf}
\usepackage{amsmath}
\usepackage{multirow}
\usepackage{xcolor}
\usepackage{ulem}
\usepackage[version=4]{mhchem}
\newcommand{\RNum}[1]{\uppercase\expandafter{\romannumeral #1\relax}}

\begin{document}

\title {Superconductivity in nickelate and cuprate superconductors with strong bilayer coupling}

\author{Zhen Fan}
\altaffiliation{The first three authors contributed equally to this work.}
\affiliation{Institute of Physics, Chinese Academy of Sciences, P.O.~Box 603, Beijing 100190, China}
\affiliation{School of Physical Sciences, University of Chinese Academy of Sciences, Beijing 100049, China}

\author{Jian-Feng Zhang}
\altaffiliation{The first three authors contributed equally to this work.}
\affiliation{Institute of Physics, Chinese Academy of Sciences, P.O.~Box 603, Beijing 100190, China}

\author{Bo Zhan}
\altaffiliation{The first three authors contributed equally to this work.}
\affiliation{Institute of Physics, Chinese Academy of Sciences, P.O.~Box 603, Beijing 100190, China}
\affiliation{School of Physical Sciences, University of Chinese Academy of Sciences, Beijing 100049, China}
\affiliation{ByteDance Research, Zhonghang Plaza, No.~43, North 3rd Ring West Road, Haidian District, Beijing 100089, China}

\author{Dingshun Lv}
\affiliation{ByteDance Research, Zhonghang Plaza, No.~43, North 3rd Ring West Road, Haidian District, Beijing 100089, China}

\author{Xing-Yu Jiang}
\affiliation{Institute of Physics, Chinese Academy of Sciences, P.O.~Box 603, Beijing 100190, China}

\author{Bruce Normand}\email{bruce.normand@psi.ch}
\affiliation{Laboratory for Theoretical and Computational Physics, Paul Scherrer Institute, CH-5232 Villigen-PSI, Switzerland}
\affiliation{Institute of Physics, Ecole Polytechnique F\'ed\'erale de Lausanne (EPFL), CH-1015 Lausanne, Switzerland}

\author{Tao Xiang}\email{txiang@iphy.ac.cn}
\affiliation{Institute of Physics, Chinese Academy of Sciences, P.O.~Box 603, Beijing 100190, China}
\affiliation{Collaborative Innovation Center of Quantum Matter, Beijing 100190, China}
\affiliation{Beijing Academy of Quantum Information Sciences, Beijing, 100190, China.}

\date{\today}

\begin{abstract}
 The discovery of superconductivity at 80 K under high pressure in \ce{La3Ni2O7} presents the groundbreaking confirmation that high-$T_c$ superconductivity is a property of strongly correlated materials beyond cuprates. We use density functional theory (DFT) calculations of the band structure of \ce{La3Ni2O7} under pressure to verify that the low-energy bands are composed almost exclusively of Ni 3$d_{x^2-y^2}$ and O 2$p$ orbitals. We deduce that the Ni 3$d_{z^2}$ orbitals are essentially decoupled by the geometry of the high-pressure structure and by the effect of the Ni Hund coupling being strongly suppressed, which results from the enhanced interlayer antiferromagnetic interaction between 3$d_{z^2}$ orbitals and the strong intralayer hybridization of the 3$d_{x^2-y^2}$ orbitals with O 2$p$. By introducing a tight-binding model for the Fermi surfaces and low-energy dispersions, we arrive at a bilayer $t$-$t_\perp$-$J$ model with strong interlayer hopping, which we show is a framework unifying \ce{La3Ni2O7} with cuprate materials possessing similar band structures, particularly the compounds \ce{La2CaCu2O6}, \ce{Pb2Sr2YCu3O8}, and \ce{EuSr2Cu2NbO8}. We use a renormalized mean-field theory to show that these systems should have ($d$+$is$)-wave superconductivity, with a dominant $d$-wave component and the high $T_c$ driven by the near-optimally doped $\beta$ band, while the $\alpha$ band adds an $s$-wave component that should lead to clear experimental signatures. 
\end{abstract}


\maketitle

\section{Introduction}

Ever since the seminal discovery of high-temperature superconductivity in copper oxide materials, scientists have explored the possibility of achieving similar transition temperatures ($T_c$) in other correlated electronic systems. Particular targets in this search include transition-metal compounds and especially oxides with crystal structures resembling those of the cuprates. In this context, the recent identification of superconductivity at $T_c \approx 80$ K in \ce{La3Ni2O7} under high pressure \cite{Sun2023SignaturesOS, hou2023emergence, zhang2023hightemperature, liu2023electronic} constitutes the long-sought demonstration of cuprate-type $T_c$ values in a non-cuprate system, and as such has attracted intensive research activity within the condensed-matter community.

At ambient pressure \ce{La3Ni2O7} is a metal. Following the discovery of high-pressure superconductivity, the band structure and Fermi surface at ambient pressure were characterized by angle-resolved photoemission spectroscopy~(ARPES) \cite{yang2023orbitaldependent}. Ongoing studies of a low-temperature instability in this metallic state currently differ in their conclusions, with resonant inelastic X-ray scattering~(RIXS) data \cite{chen2024electronic} being interpreted as ($\pi/2$,$\pi/2$,$\pi$) spin-density-wave order below 150 K and neutron scattering data \cite{xie2024neutron} indicating strong antiferromagnetic (AF) spin fluctuations but no order above 10 K. When pressure is applied, the crystal structure of \ce{La3Ni2O7} undergoes a phase transition from the orthorhombic symmetry $Amam$ to $Fmmm$ at approximately 14 GPa, where superconductivity emerges \cite{Sun2023SignaturesOS, liu2023electronic, hou2023emergence, zhang2023hightemperature}. This change of crystal structure calls into question whether the ambient-pressure magnetic state can in any sense be regarded as a ``parent phase'' of the superconducting state. 

Turning to the superconductivity, measurements made at 20.5 GPa showed that, similar to the cuprates, \ce{La3Ni2O7} exhibits a resistivity that is linear in temperature above $T_c$~\cite{zhang2023hightemperature}, reflecting the importance of strong electronic correlations. Measurements of the Hall coefficient showed further that the transport is dominated by holes~\cite{2023Zhou}. However, the superconducting volume fraction, deduced from measurements of the modulated a.c.~magnetic susceptibility, was found to be very low (in the 1-5\% range), and obtaining the zero-resistance state was difficult in many samples \cite{zhang2023hightemperature, 2023Zhou, puphal2023unconventional, dong2023visualization}, implying a high degree of inhomogeneity. While some authors suggested the role of O nonstoichiometry, other work identified an unconventional monolayer-trilayer~(1313) phase in single crystals grown in an optical floating-zone furnace \cite{puphal2023unconventional}; although the 1313 phase showed a resistance drop at high pressure similar to the onset of superconductivity, zero resistance and diamagnetism have not been observed. Very recent studies seeking to alter the crystal-growth protocol appear to have achieved more robust superconductivity with volume fractions close to 50\% \cite{wang2023observation, jcpc, li2024pressuredriven}, reinforcing the suggestion that the fragility of superconductivity in \ce{La3Ni2O7} is a consequence of inhomogeneous crystal structures; from the differing results found at ambient pressure (previous paragraph), one may deduce that the problem of sample inhomogeneity is not restricted to the high-pressure regime. We continue to assume that the high-$T_c$ superconducting phase is a property of the bilayer (2222) structure and focus on this henceforth. Because it is not possible to probe the band structure under high pressure by ARPES, all the preliminary insight into the origin of superconductivity in \ce{La3Ni2O7} has to date been provided by density functional theory~(DFT) calculations~\cite{PhysRevLett.131.126001, gu2023effective, zhang2023electronic, sakakibara2023possible, lechermann2023electronic, yang2023minimal, PhysRevLett.131.206501, PhysRevB.108.125105,cao2023flat}.

DFT results representative of the bilayer $Fmmm$ structure reveal that three electronic bands, designated as $\alpha$, $\beta$, and $\gamma$, cross the Fermi level. The $\alpha$ band forms an electron pocket, while the $\beta$ and $\gamma$ bands create two distinct hole pockets in the first Brillouin zone. Considering the two relevant $d$ orbitals of the Ni atom, the $\alpha$ and $\beta$ bands originate primarily from the Ni 3$d_{x^2-y^2}$ and O 2$p_{x,y}$ orbitals, whereas the $\gamma$ band arises predominantly from the Ni 3$d_{z^2}$ orbital. Although DFT methods provide only an approximate treatment of strongly correlated materials, they can be calibrated against ARPES measurements at ambient pressure, which indicates that the leading correlation effects in \ce{La3Ni2O7} are being captured appropriately. At this pressure, while the $\alpha$ and $\beta$ pockets exhibit shapes very similar to those in the high-pressure system, the $\gamma$ band lies below the Fermi surface.

A key difference between \ce{La3Ni2O7} and the cuprate superconductors lies in the valence-electron numbers of the Ni$^{2+}$ and Cu$^{2+}$ ions. Specifically, a Cu ion ($3d^9$) possesses only one unpaired valence electron in the 3$d_{x^2-y^2}$ orbital, while the Ni ion (conventionally $3d^8$) has unpaired valence electrons in both the 3$d_{x^2-y^2}$ and 3$d_{z^2}$ orbitals. Multiple models have already been proposed that incorporate different Hubbard- or $t$-$J$-type interactions in a bid to capture the properties of this multi-orbital system \cite{PhysRevB.108.L140505, tian2023correlation, lechermann2023electronic, liu2023spmwave, liao2023electron, kaneko2023pair, ryee2023critical}. Most of these models consider interactions within the 3$d_{x^2-y^2}$ and 3$d_{z^2}$ orbitals, deducing both an intralayer AF interaction between the 3$d_{x^2-y^2}$ orbitals and an interlayer AF interaction between the 3$d_{z^2}$ orbitals \cite{gu2023effective, wu2023charge, PhysRevB.108.174501, luo2023hightc, schlomer2023superconductivity, chen2023orbitalselective, qu2023roles, oh2023type, jiang2023high}. In contrast, other models focus exclusively on the latter (interlayer) interaction \cite{shen2023effective, yang2023minimal, qin2023hightc}. The Hund coupling, $J_{\rm H}$, between the 3$d_{x^2-y^2}$ and 3$d_{z^2}$ orbitals is considered explicitly in certain two-orbital models \cite{luo2023hightc, qu2023roles}, and in the limit of strong $J_{\rm H}$ it has been argued that the interlayer interaction between the 3$d_{z^2}$ orbitals may induce an effective interlayer AF interaction between 3$d_{x^2-y^2}$ spins. This leads to a one-orbital $t$-$J$-$J_\perp$ model with bilayer AF interactions \cite{lu2023interlayer, oh2023type, qu2023bilayer, schlomer2023superconductivity, Ludachuan2023, zhang2023strong}.

A key point of contention among these models lies in the interaction terms. The focus of many studies on the interlayer interaction between 3$d_{z^2}$ orbitals is justified by the fact that the interlayer Ni-O-Ni angle jumps from $168^\circ$ to $180^\circ$ at the 14 GPa structural transition, while the $c$-axis lattice constant is shortened. Both changes should contribute to a significant enhancement of the interlayer interaction at pressures $P > 14$ GPa, leading to a popular interpretation of the high-pressure superconductivity in \ce{La3Ni2O7} involving enhanced interaction-induced interlayer pairing between electrons in 3$d_{z^2}$ orbitals \cite{gu2023effective, liu2023spmwave, wu2023charge, lu2023interlayer, shen2023effective, qu2023bilayer, lange2023pairing, yang2023minimal, qin2023hightc, sakakibara2023possible, Ludachuan2023, PhysRevB.108.174501,
luo2023hightc, schlomer2023superconductivity}. Within this class of models, there are two proposed mechanisms for the origin of pairing, one involving the metallization of the $\gamma$ band when the high pressure lifts it through the Fermi surface \cite{2015GaoLuXiang, Sun2023SignaturesOS, shen2023effective}. The other relies on strong hybridization between the 3$d_{z^2}$ and 3$d_{x^2-y^2}$ orbitals, with the latter contributing significantly to the propagation of paired electrons, whose pairing is induced predominantly by the interactions between the Ni 3$d_{z^2}$ orbitals) \cite{gu2023effective,liu2023spmwave,wu2023charge,lu2023interlayer, qu2023bilayer, lange2023pairing,yang2023minimal,qin2023hightc}. In counterpoint to these $d$-orbital models, another strand of research has considered the role of the O 2$p$ orbitals \cite{Sun2023SignaturesOS, PhysRevLett.131.126001, wu2023charge, lechermann2023electronic,dong2023visualization}, whose strong hybridization with the 3$d_{x^2-y^2}$ orbitals would suggest the emergence of Zhang-Rice singlets \cite{Sun2023SignaturesOS, wu2023charge, PhysRevLett.131.126001,jiang2023high}. 

In the broadest terms, these models for the superconducting properties of \ce{La3Ni2O7} under high pressure can be classified into two groups. The first assumes that the Ni 3$d_{z^2}$ band emerging above the Fermi surface is the driver of high-$T_c$ superconductivity, through the interlayer Ni 3$d_{z^2}$ antiferromagnetic interaction. The second assumes that the superconducting pairing is driven by the interplay between Ni $d_{x^2-y^2}$ and O 2$p_{x,y}$ orbitals, but differs from the cuprates by a strong interlayer antiferromagnetic interaction between Ni $d_{x^2-y^2}$ electrons induced by the strong Hund coupling. Exploration of the pairing symmetry within these different classes of model has also yielded diverse results, including different types of $s$-wave \cite{yang2023minimal, gu2023effective, liu2023spmwave, oh2023type, qu2023bilayer, zhang2023trends, tian2023correlation, PhysRevB.108.L140505, zhang2023structural, PhysRevB.108.174501, Ludachuan2023, sakakibara2023possible, qin2023hightc, tian2023correlation}, $d$-wave \cite{lechermann2023electronic, jiang2023high}, and more complex cases~\cite{lu2023interlayer, liao2023electron, luo2023hightc, ryee2023critical, oh2023type, jiang2023pressure}. Most of these theoretical investigations nevertheless either address superconductivity in \ce{La3Ni2O7} independently or tend to emphasize the differences between \ce{La3Ni2O7} and the cuprates, rather than any common ingredients. 

\begin{figure*}[t]
\includegraphics[width=0.9\linewidth]{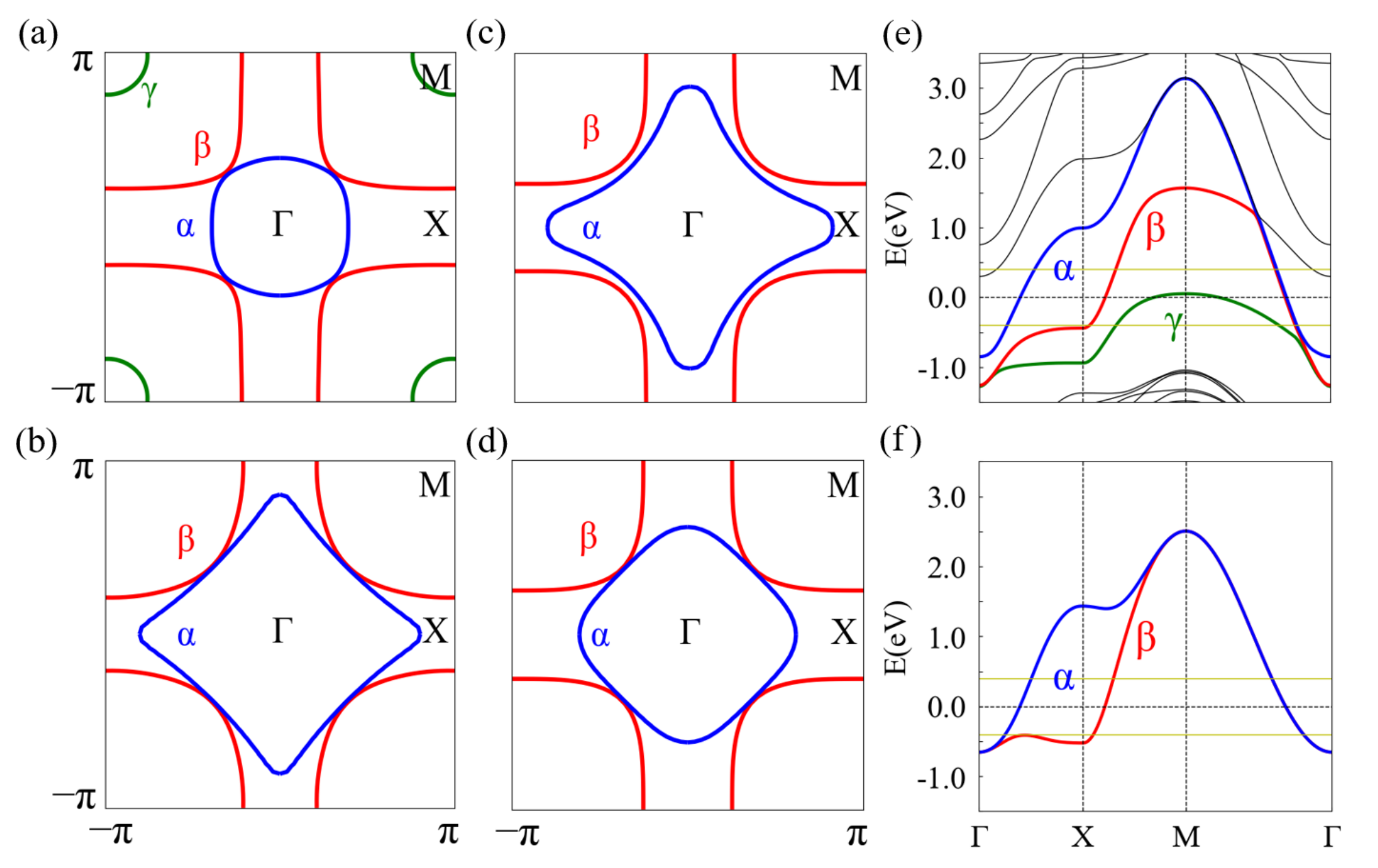}
\caption{Fermi surfaces at $k_z = 0$ calculated by DFT for (a) \ce{La3Ni2O7} under high pressure, and for three cuprates at ambient pressure, \ce{La2CaCu2O6} (b), \ce{Pb2Sr2YCu3O8} (c), and \ce{EuSr2Cu2NbO8} (d). (e) Low-energy band dispersions calculated by DFT for \ce{La3Ni2O7} under high pressure, shown along the high-symmetry lines in the two-dimensional (2D) Brillouin zone. (f) Low-energy band dispersions described by the effective bilayer model of Eq.~(\ref{eq:2tb}) for the $\alpha$ and $\beta$ bands. The parameters were determined by fitting to the DFT results of the panel (e) in the energy interval [$-0.4$ eV, 0.4 eV] around the Fermi level.}
\label{fig:cula}
\end{figure*}

In this work, we reconsider the fundamental factors governing the superconducting properties of \ce{La3Ni2O7} and establish a direct connection with cuprate materials. The crucial insight arises from examining the Fermi surface of \ce{La3Ni2O7} under high pressure, determined by DFT calculations as in Refs.~\cite{PhysRevLett.131.126001, gu2023effective, zhang2023electronic, sakakibara2023possible, lechermann2023electronic, yang2023minimal}. We observe striking similarities between the Fermi surfaces of \ce{La3Ni2O7} and those of a number of bilayer cuprates with medium-high $T_c$ values, including \ce{La2CaCu2O6} \cite{CAVA1990138, Cava1990, ANSALDO19911213}, \ce{Pb2Sr2YCu3O8} \cite{KRESS19891345, ZANDBERGEN198981}, and \ce{EuSr2Cu2NbO8} \cite{133392, KIM2013165}. As summarized in Ref.~\cite{PhysRevB.89.224505}, the bilayer splitting is larger in these three systems than in the better-known cuprate superconductors, presumably because in-plane buckling is larger and the energy difference between the two $e_g$ orbitals is smaller, and this causes the similarity to \ce{La3Ni2O7}. In Fig.~\ref{fig:cula} we show how the Fermi surfaces of the $\alpha$ and $\beta$ bands in \ce{La3Ni2O7} closely mirror those of the three cuprate superconductors listed. The sole distinction lies in the fact that \ce{La3Ni2O7} possesses an additional $\gamma$ Fermi surface under high pressure. Given the significant impact of the Fermi-surface shape on the low-energy physics, this unexpected resemblance prompts a reevaluation of the roles played by the three bands in \ce{La3Ni2O7}. As a consequence, we center our investigation on identifying the common elements shared by these materials, with the overarching goal of proposing a universal description of superconductivity applicable both to \ce{La3Ni2O7} and to the cuprates.

The model we deduce is then different from both of the model classes summarized above. Although we conclude that the superconducting pairing is driven by the Ni 3$d_{x^2-y^2}$ and O 2$p_{x,y}$ orbitals, unlike the second class of models we find that superconductivity is driven not by the interlayer antiferromagnetic interaction induced by the Hund coupling, but by intralayer antiferromagnetic interactions as in the cuprates. For us the Ni 3$d_{z^2}$ bands are neither driving nor driven by the superconductivity, and act only as an electron pool that modulates the hole doping of the superconducting 3$d_{x^2-y^2}$ band. To show all the details that lead to these conclusions, the subsequent sections of this paper are organized as follows. In Sec.~\ref{Sec:Model}, we analyze the low-energy electronic structure of \ce{La3Ni2O7}, which leads us to put forward a bilayer $t$-$t_\perp$-$J$ model. In Sec.~\ref{phase}, we use a renormalized mean-field theory to calculate the superconducting phase diagram of this model, focusing on the symmetry of the pairing gap and the effect of doping. Section \ref{sum} provides a summary of our results and of their physical implications for nickelate superconductivity.

\section{Minimal model for L\MakeLowercase{a}\textsubscript{3}N\MakeLowercase{i}\textsubscript{2}O\textsubscript{7}} 
\label{Sec:Model}
 
To identify the key interactions driving the high-$T_c$ superconducting pairing in bilayer \ce{La3Ni2O7} under pressure, it is essential to understand two primary physical effects. The first involves an accurate treatment of the interlayer coupling and the second centers on the role of the Hund coupling, $J_{\rm H}$. A noteworthy characteristic of the Fermi surfaces in \ce{La3Ni2O7} is the degeneracy of the $\alpha$ and $\beta$ bands along the two diagonal directions in the Brillouin zone. As we discuss in Sec.~\ref{Bisp}, this degeneracy is the hallmark of a special form of interlayer tunneling within the bilayer, which forms a critical piece of the puzzle when constructing a minimal low-energy model. Hund coupling is inherent to the Ni ion but, as we detail in Sec.~\ref{SubSec:t-J}, its effects are suppressed quite considerably at high pressures, where the Ni 3$d_{z^2}$ orbitals on adjacent layers experience an enhanced interaction. At the same time, the hybridization of the Ni 3$d_{x^2-y^2}$ orbitals with the O $p$ orbitals, where the holes reside, acts to create both strong intralayer AF interactions and Zhang-Rice singlets, which further reduce the relevance of the Hund term. Drawing from these analyses, we deduce that the minimal model for the low-energy physics of \ce{La3Ni2O7} is the single-band bilayer $t$-$t_\perp$-$J$ model with strong interlayer hopping, $t_\perp$, and hence that the high-$T_c$ pairing has the same origin as in cuprate superconductors.

\subsection{Bilayer coupling in L\MakeLowercase{a}\textsubscript{3}N\MakeLowercase{i}\textsubscript{2}O\textsubscript{7}}
\label{Bisp}

We have performed our own DFT calculations of the electronic band structures of \ce{La3Ni2O7} in its high-pressure $Fmmm$ phase and of the three cuprates \ce{La2CaCu2O6}, \ce{Pb2Sr2YCu3O8}, and \ce{EuSr2Cu2NbO8}, including correlation effects by standard methodology. The technical details of these calculations are presented in App.~A and the results shown in Figs.~\ref{fig:cula}(a-e) reproduce well the results reported by other authors both for \ce{La3Ni2O7} at the experimentally characterized \cite{Sun2023SignaturesOS} pressure of 29.5 GPa \cite{PhysRevLett.131.126001, gu2023effective, zhang2023electronic, sakakibara2023possible, lechermann2023electronic, yang2023minimal, PhysRevLett.131.206501, PhysRevB.108.125105, cao2023flat} and for the three cuprates \cite{PhysRevB.89.224505}. All of these calculations find that the $\gamma$ band has almost exclusively Ni 3$d_{z^2}$ orbital character at the Fermi surface while the $\alpha$ and $\beta$ bands are predominantly of 3$d_{x^2-y^2}$ nature, with a 3$d_{z^2}$ admixture that averages to approximately 20\% around the Fermi surface, varying quite uniformly from 40\% in the $k_x$ and $k_y$ directions in the Brillouin zone to 0 when $k_x = \pm k_y$. We stress here that the difference between our work and the work of other groups performing these calculations lies not at the level of DFT but at the level of interpretation. Starting with the same complete set of orbitals and interactions, we apply the logic of understanding the two physical effects described above to deduce a rather different minimal model.  

We begin by noting that a distinctive feature of the $\alpha$- and $\beta$-bands, which recurs in all these DFT calculations, is their degeneracy along the two diagonal directions, which is shown at the Fermi surface in Figs.~\ref{fig:cula}(a,b,d) and for part of the $\Gamma$M direction in Fig.~\ref{fig:cula}(e). This phenomenon is well known in bilayer high-$T_c$ cuprate superconductors \cite{PhysRevB.89.224505}, having been identified in \ce{YBa2Cu3O7}$_{-\delta}$, \ce{Ba2Sr2CaCu2O8}, \ce{HgBa2CaCu2O6}, and \ce{Tl2Ba2CaCu2O8}, as well as in two of the three cuprates whose Fermi surfaces we show in Fig.~\ref{fig:cula}. It was recognized very early that the cuprates are charge-transfer insulators, with the doped holes residing on the O 2$p_x$ or 2$p_y$ orbitals and forming Zhang-Rice singlets~\cite{PhysRevB.37.3759} with the unpaired electron in a Cu 3$d_{x^2-y^2}$ orbital due to their strong $\sigma$-hybridization. The wave function of the O hole in this singlet state adopts the 3$d_{x^2-y^2}$ symmetry of the Cu orbital and hence the interlayer tunneling of an O hole depends on its in-plane momentum, $(k_x, k_y)$, and vanishes when $k_x = \pm k_y$ \cite{1996Xiang, 1998Xiang}, leading to the degeneracy of the bilayer bands along the Brillouin-zone diagonals. It should be emphasized that direct hopping between Cu 3$d_{x^2-y^2}$ orbitals along the $z$-axis is negligible due to the vanishing wave-function overlap in this direction, and the interlayer tunneling is facilitated by weak admixtures with the Cu 3$d_{z^2}$ and 4$s$ orbitals, which are rotation-symmetric around the $z$-axis. 

The $\alpha$ and $\beta$ bands in \ce{La3Ni2O7} are a simple consequence of strong bilayer splitting, which completely alters the curvature of the $\alpha$ Fermi surface. Although this splitting is stronger than in the most famous examples of high-$T_c$ cuprate superconductors, namely \ce{YBa2Cu3O7}$_{-\delta}$, \ce{Ba2Sr2CaCu2O8}, \ce{HgBa2CaCu2O6}, and \ce{Tl2Ba2CaCu2O8} as listed above, it is completely in line with the splittings, and hence the Fermi-surface shapes, found in \ce{La2CaCu2O6}, \ce{Pb2Sr2YCu3O8}, and \ce{EuSr2Cu2NbO8}, as we show in Figs.~\ref{fig:cula}(b-d). Although the degeneracy is partially broken in \ce{Pb2Sr2YCu3O8}, the fundamental features of the Fermi surfaces remain unaltered. We note here that these three cuprates do not share quite the same high $T_c$ values as their better-known analogs; at hole doping $\delta = 0.2$, the superconducting transition temperatures are approximately 60 K for both \ce{La2CaCu2O6} \cite{Cava1990} and \ce{Pb2Sr2YCu3O8} \cite{ZANDBERGEN198981}, and 40 K for \ce{EuSr2Cu2NbO8} \cite{KIM2013165}. 

As noted above, in our calculations we also find that the $\alpha$ and $\beta$ bands at $k_x = \pm k_y$ are formed almost exclusively from the electronic contributions of the Ni 3$d_{x^2-y^2}$ and in-plane O 2$p_x$ and 2$p_y$ orbitals. This observation implies that the nodal degeneracy observed in \ce{La3Ni2O7} is unrelated to the Ni 3$d_{z^2}$ and apical O 2$p_z$ orbitals. Indeed, the fact that the energy dispersions of the $\alpha$ and $\beta$ bands in \ce{La3Ni2O7}, and the shapes of their Fermi surfaces, are so directly comparable to those in \ce{La2CaCu2O6}, \ce{Pb2Sr2YCu3O8}, and \ce{EuSr2Cu2NbO8}, strongly suggests a common origin for the high-$T_c$ pairing in these materials, encoded within the $\alpha$ and $\beta$ bands alone. 

Here we comment that, although the 3$d_{z^2}$ orbitals make a negligible contribution to the $\alpha$ and $\beta$ bands around the Fermi level, they do have a finite involvement arising from hybridization with the in-plane O 2$p$ orbitals. This contribution has a notable effect on the high-energy states in both bands, breaking the degeneracy between the $\alpha$ and $\beta$ bands along the two zone-diagonal directions, as shown in Fig.~\ref{fig:cula}(e). If this additional hybridization is neglected in a minimal low-energy model, the two bands remain degenerate along the entirety of the zone diagonals, as shown in Fig.~\ref{fig:cula}(f).

\subsection{Effective bilayer $t$-$t_\perp$-$J$ model} 
\label{SubSec:t-J}

 The above discussion implies that the electronic states around the Fermi level in the $\alpha$ and $\beta$ bands can be described effectively by a bilayer model of Ni 3$d_{x^2-y^2}$ electrons coupled through an interlayer tunneling matrix element $t_{\perp\mathbf{k}}$. This model is expressed by the Hamiltonian
\begin{equation}
    H_0 = \sum_{\mathbf{k},\sigma}
    \left(
    \begin{array}{cc}
        c^\dagger_{1,\mathbf{k}} & c^\dagger_{2,\mathbf{k}}
    \end{array}
    \right)
    \left(
     \begin{array}{cc}
        \varepsilon_\mathbf{k} & t_{\perp \mathbf{k} }\\
        t_{\perp \mathbf{k} } & \varepsilon_\mathbf{k}
    \end{array}
    \right)
    \left(
    \begin{array}{c}
        c_{1,\mathbf{k}} \\
        c_{2,\mathbf{k}}
    \end{array}
    \right) ,
    \label{eq:2tb}
\end{equation}
where $c_{n,\mathbf{k}}$ is the annihilation operator of a 3$d_{x^2-y^2}$ electron in layer $n = 1$, 2 and
\begin{eqnarray}
    \varepsilon_\mathbf{k} & = & - 2t\left(\cos k_x - \cos k_y\right) - 4t^\prime \cos k_x \cos k_y \nonumber \\
    && - 2t^{\prime\prime} (\cos 2k_x - \cos 2k_y )  - \mu
\end{eqnarray}
is the dispersion relation of electrons in each layer, with respective hopping integrals $t$, $t^\prime$, and $t^{\prime\prime}$ between two nearest-, next-, and next-next-neighbor sites; $\mu$ is the chemical potential.

In Eq.~(\ref{eq:2tb}), $ t_{\perp \mathbf{k}}$ is the interlayer hopping term,
\begin{eqnarray}
    t_{\perp \mathbf{k}} = 4t_\perp \! \left[ \cos k_x - \cos k_y + \eta (\cos 2k_x - \cos 2k_y )\right]^2 \! , \;\;
\label{tperp}
\end{eqnarray}
whose form arises from the hopping of an orbital possessing 3$d_{x^2-y^2}$ symmetry and hence contains only ($\cos k_x - \cos k_y$) and its higher harmonics. Complementing the ($\cos k_x - \cos k_y$) term used conventionally to describe the $c$-axis hopping, we introduce the $\eta$ term to enhance the accuracy of the fit to the DFT bands.
 
To obtain a tight-binding fit to the $\alpha$ and $\beta$ bands obtained from DFT, shown for high-pressure \ce{La3Ni2O7} in Fig.~\ref{fig:cula}(e), we focus on the energy range from $- 0.4$ eV to 0.4 eV. By fitting the eigenenergies obtained from the Hamiltonian of Eq.~(\ref{eq:2tb}), we determine the parameter values for a pressure equivalent to 30 GPa as $t = 0.39$ eV, $t^\prime/t = - 0.18$, $t^{\prime\prime}/t = 0.05$, $t_\perp/t = - 0.14$, $\eta = 0.15 $, and $\mu = - 0.73$ eV. These parameters provide a very accurate fit to the Fermi surface of \ce{La3Ni2O7} [Fig.~\ref{fig:cula}(a)], as shown in App.~A. We draw attention to the fact that $t$ has quantitatively the same magnitude as in the well-known bilayer high-$T_c$ cuprates; $t^\prime$ and $t^{\prime\prime}$ naturally have the same relation to $t$, required to generate the familiar hole-like Fermi surface, while $t_\perp$ is four times larger than the value used to model \ce{YBa2Cu3O7}$_{-\delta}$ and \ce{Ba2Sr2CaCu2O8} in early literature \cite{rnkf}. We comment again that these effective values are optimized to reproduce the low-energy electronic structure of the $\alpha$ and $\beta$ bands only, and deviate significantly from their band structures beyond the energy range of the fit [Fig.~\ref{fig:cula}(f)]; as noted above, the primary reason for this deviation is the increasing Ni 3$d_{z^2}$ orbital content that sets in at energies of order 2 eV. 

Turning from the $t$ term to the $J$ term, a notable complication in constructing a minimal model for the interactions in \ce{La3Ni2O7} is the Hund coupling between the Ni 3$d_{z^2}$ and 3$d_{x^2-y^2}$ orbitals. However, as we summarized at the beginning of this section, two effects conspire to quench the effect of the Hund coupling in the superconducting phase of the bilayer nickelate. The first effect is the pressure-induced structural transition occurring at 14 GPa in \ce{La3Ni2O7}, which changes the $c$-axis Ni–O–Ni bond angle from 168$^\circ$ to precisely 180$^\circ$ while also shortening the bond lengths, and thus enhancing significantly the superexchange interaction between the two Ni 3$d_{z^2}$ orbitals in adjacent layers of a bilayer. As detailed in Ref.~\cite{jiang2023pressure}, the strengthening of this AF interaction causes a concomitant suppression of the effect of the Hund coupling of the 3$d_{z^2}$ to the 3$d_{x^2-y^2}$ orbitals. 

In a similar vein, the second effect is the strong hybridization of the Ni 3$d_{x^2-y^2}$ orbitals with the in-plane O $p$ orbitals. In the absence of doped holes on the O ions, this gives rise to a robust intralayer AF coupling of Ni 3$d_{x^2-y^2}$ spins, and in the presence of a doped hole it causes the formation of a Zhang-Rice singlet, which quenches the Ni 3$d_{x^2-y^2}$ spin. Both situations constitute the dominant correlations of the 3$d_{x^2-y^2}$ orbitals, contributing again to making their Hund coupling to the 3$d_{z^2}$ orbitals ineffective. Although this hybridization is further enhanced in the high-pressure $Fmmm$ phase, where the NiO$_2$ planes become significantly flatter, it is generically strong in the $\sigma$-bonding geometry. Given the nontrivial valence state of a Ni atom in \ce{La3Ni2O7}, which diverges from the straightforward $+2$ configuration, we make the case that the two $e_g$ orbital types are effectively decoupled into separate subsystems, visible in the quite different energies of their respective bands [Fig.~\ref{fig:cula}(e-f)], and this allows them to be treated independently for the purpose of revealing the low-energy physics. 

This discussion leads to the conclusion that the 3$d_{z^2}$ orbitals are not relevant to the superconducting properties of \ce{La3Ni2O7}, which remain governed by the Zhang-Rice physics of the 3$d_{x^2-y^2}$ orbitals familiar from the cuprate superconductors. The minimal model for the interaction in this situation is a regular in-plane superexchange term, leading to the Hamiltonian 
\begin{equation}
    H = H_0 + J \sum_{n \langle ij \rangle} \left(\mathbf{S}_{ni} \cdot \mathbf{S}_{nj} - {\textstyle{\frac{1}{4}}} n_{ni} n_{nj} \right) ,
    \label{eq:tJ_model}
\end{equation}
where $\mathbf{S}_{ni}$ denotes the spin operator of the electron in the 3$d_{x^2-y^2}$ orbital at site $i$ in layer $n$ of the bilayer. This Hamiltonian is defined within the subspace obeying the constraint of no double occupancy at each lattice site. Thus we arrive at a standard one-orbital bilayer $t$-$t_\perp$-$J$ model, albeit with a sufficiently strong interlayer hopping, $t_{\perp}$, that the $\alpha$ band is converted from hole-like to electron-like.

The model we have deduced differs from all of the previous proposals made for \ce{La3Ni2O7} in two aspects. First, the pairing interaction is not a consequence of the AF interlayer interaction between 3$d_{z^2}$ spins but instead is governed by the AF intralayer interaction between 3$d_{x^2-y^2}$ spins. Second, the interlayer hopping term for the 3$d_{x^2-y^2}$ holes is strong and anisotropic in ${\bf k}$, and drives the emergence of the characteristically different $\alpha$ and $\beta$ bands. This significant bilayer-splitting effect has been overlooked in most studies to date.

\section{Superconducting phase diagram} 
\label{phase}

An analytical treatment of the $t$-$J$ model is made extremely difficult by the hard local constraint of no double site occupancy. In Ref.~\cite{FCZhang_1988}, Zhang and coworkers introduced a renormalized mean-field theory that aimed to capture the fundamental consequences of the constrained (three-state) site basis at the lowest order as a renormalization of both the hopping and the interaction terms. In a one-orbital model on a square lattice with only a dilute concentration, $\delta$, of doped holes, the renormalized coefficients are 
\begin{eqnarray}
     t_\mathrm{eff} & = & \frac{2 \delta}{1 + \delta} t , \label{eq:teff} \\
     J_\mathrm{eff} & = & \frac{4}{(1 + \delta)^2} J , \label{eq:jeff}
\end{eqnarray}
(and similarly for the other hopping integrals). These expressions contain a strong reduction of the effective bandwidth of the mobile holes, arising from the difficulty of propagating in an AF background, and a strong enhancement of the spin correlations arising from the increased weight of the 
contributing configuration in the constrained manifold. To gain insight into the physics of the bilayer $t$-$J$ model for \ce{La3Ni2O7} under pressure and its analogous cuprates, we apply this renormalized mean-field theory by replacing the coupling constants in Eq.~(\ref{eq:tJ_model}) with their renormalized counterparts, thereby releasing the constraint of no double occupancy at each lattice site.

In the bilayer system with strong $t_\perp$, we observe a large difference between the electron occupancies, $n_\alpha$ and $n_\beta$, of the $\alpha$ and $\beta$ bands [Figs.~\ref{fig:cula}(a-d)]. When $t_\perp$ is rather weak, which is the high-$T_c$ cuprate paradigm realized in \ce{YBa2Cu3O7}$_{-\delta}$, \ce{Ba2Sr2CaCu2O8}, and similar materials, then the hole dopings $\delta_\alpha = 1 - n_\alpha$ and $\delta_\beta = 1 - n_\beta$ can both remain in the range around optimal doping. With strong $t_\perp$, however, one of the bands becomes strongly overdoped and the other strongly underdoped ($\delta_\beta$ may even become negative), as shown in Figs.~\ref{fig:cula}(b-d), presumably accounting for the lower $T_c$ values in the cuprates \ce{La2CaCu2O6}, \ce{Pb2Sr2YCu3O8}, and \ce{EuSr2Cu2NbO8}. In high-pressure \ce{La3Ni2O7} [Fig.~\ref{fig:cula}(a)], the lower total electron count results in $\delta_\beta = 0.15$ being close to optimal doping, whereas $\delta_\alpha \approx 0.75$ is extremely large. Because the renormalized mean-field theory \cite{FCZhang_1988} is formulated in real space, the relevant doping parameter is $\delta = {\textstyle \frac12} (\delta_\alpha + \delta_\beta)$, which due to the large value of $\delta_\alpha$ lies in the range 30-50\%. While one may question the applicability of the renormalized mean-field framework of Eqs.~(\ref{eq:teff}-\ref{eq:jeff}) at such large $\delta$ values, we do not enter into these considerations for our present qualitative purposes; we note only that both renormalization factors become rather small (of order 2) and then the theory approaches, but presumably remains superior to, the unrenormalized and unconstrained limit. 

\begin{figure}[t]
\centering
\includegraphics[width=1.0\linewidth]{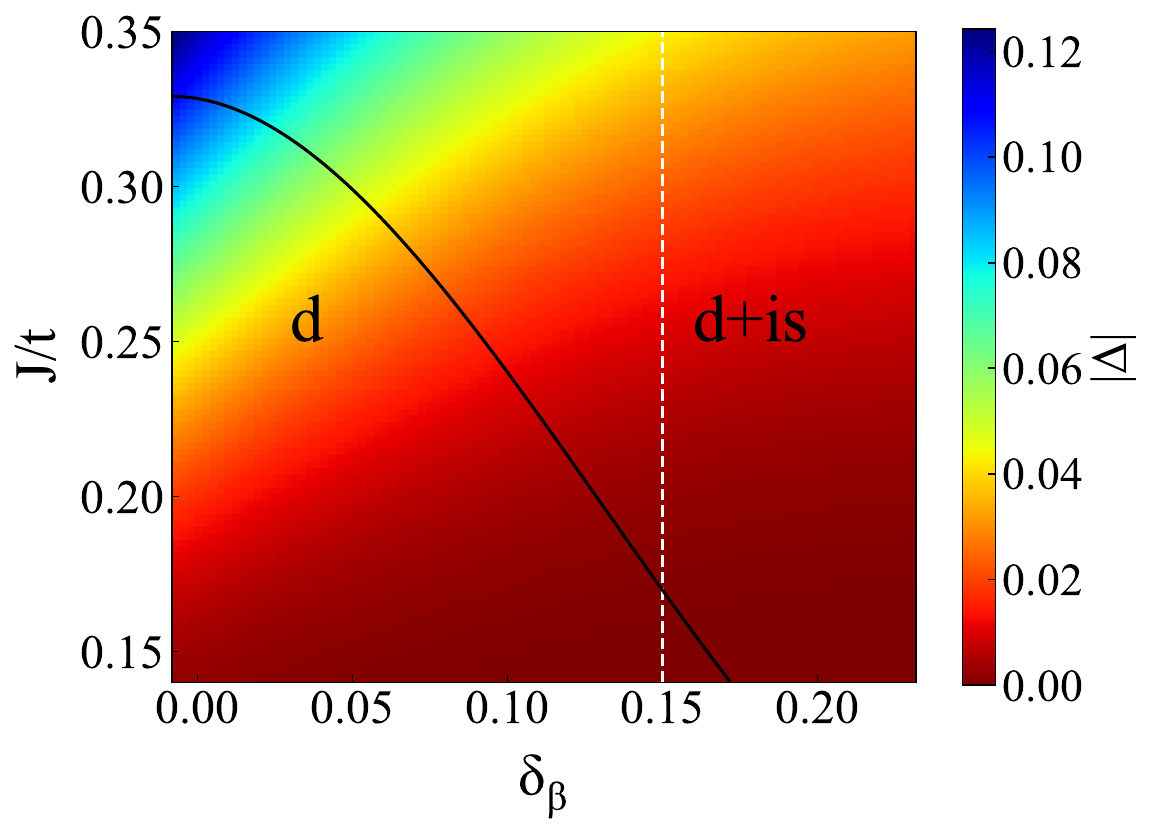}
\caption{Phase diagram of the superconducting ground state of the bilayer $t$-$J$ model deduced from the renormalized mean-field approach as a function of the $\beta$-band hole doping, $\delta_\beta$, and the interaction ratio, $J/t$. The solid black line is the phase boundary between states of $d$- and ($d$+$is$)-wave pairing. The dashed white line represents the hole doping of the $\beta$ band in \ce{La3Ni2O7} under the representative pressure of our DFT calculations [Fig.~\ref{fig:cula}(a)]. The color contours represent the magnitude of the superconducting gap in units of $t$. }
\label{fig:phase_diagram}
\end{figure}

To apply the renormalized mean-field theory, we decouple the interaction term of Eq.~(\ref{eq:tJ_model}) into the pairing channel using the mean-field approximation
\begin{eqnarray}
    2 \mathbf{S}_{ni} \cdot \mathbf{S}_{nj} & - & {\textstyle \frac12} n_{ni} n_{nj} \nonumber \\
    & \approx & -  \left( \Delta_{nij} \hat\Delta_{nij}^\dagger + h.c. \right) + \left| \Delta_{nij} \right|^2,
\end{eqnarray}
where $\hat\Delta_{nij} = c_{n,i,\uparrow} c_{n,j,\downarrow} - c_{n,i,\downarrow} c_{n,j,\uparrow}$ and $\Delta_{nij}$ is the pairing order parameter in real space. In momentum space, the mean-field Hamiltonian becomes
\begin{equation}
H_{\rm MF} = \sum_{\mathbf{k}} \Psi_{\mathbf{k}}^{\dagger}
\begin{pmatrix}
\xi_{\mathbf{k}} & \Delta_{1,\mathbf{k}} & - \xi_{\perp\mathbf{k}} & 0 \\
\Delta_{1,\mathbf{k}}^{\star} & - \xi_{\mathbf{k}} & 0 & \xi_{\perp\mathbf{k}} \\
- \xi_{\perp\mathbf{k}} & 0 & \xi_{\mathbf{k}} & \Delta_{2,\mathbf{k}} \\
0 & \xi_{\perp\mathbf{k}} & \Delta_{2,\mathbf{k}}^{\star} & - \xi_{\mathbf{k}}
\end{pmatrix}
\Psi_{\mathbf{k}},
\end{equation}
where $\Psi_{\mathbf{k}}^{\dagger} = \big( c_{1,\mathbf{k},\uparrow}^{\dagger}, c_{1,-\mathbf{k},\downarrow}, c_{2,\mathbf{k},\uparrow}^{\dagger}, c_{2,-\mathbf{k},\downarrow} \big)$, $\xi_{\mathbf{k}} = t_{\delta} \varepsilon_{\mathbf{k}}$, and $\xi_{\perp\mathbf{k}} = t_{\delta} t_{\perp\mathbf{k}}$ with $t_{\delta} = 2 \delta/(1 + \delta)$. $\Delta_{n,\mathbf{k}}$ is the gap function in layer $n$ and may have different components with different pairing symmetries; anticipating that this cuprate-type problem will yield singlet superconductivity, the most general $d$+$is$-pairing state would have 
\begin{equation}
\Delta_{n,\mathbf{k}} = \Delta_{nd} \beta_{\mathbf{k}} + i \Delta_{ns} \gamma_{\mathbf{k}},
\end{equation}
where $\beta_{\mathbf{k}} = 2(\cos k_x - \cos k_y)$ and $\gamma_{\mathbf{k}} = 2 (\cos k_x + \cos k_y)$. The order parameters $\Delta_{nd}$ and $\Delta_{ns}$ are then determined from the gap equations
\begin{eqnarray}
\Delta_{nd} & = - & \frac{J_\mathrm{eff}}{2N} \sum_{\mathbf{k}} \beta_{\mathbf{k}} \langle d_{n,-\mathbf{k},\downarrow} d_{n,\mathbf{k},\uparrow} \rangle , \\
\Delta_{ns} & = & \frac{i J_\mathrm{eff}}{2N} \sum_{\mathbf{k}} \gamma_{\mathbf{k}} \langle d_{n,-\mathbf{k},\downarrow} d_{n,\mathbf{k},\uparrow} \rangle, 
\end{eqnarray}
where $N$ is the number of sites in the finite lattice on which we solve the mean-field tight-binding problem. We comment here that we did not enforce $\Delta_{1,\mathbf{k}} = \Delta_{2,\mathbf{k}}$ in order to leave open the possibility of a solution with ($d$+$is$) symmetry in one layer and ($d$$-$$is$) in the other, which would be a state preserving time-reversal symmetry, or indeed a solution with $s_{+-}$ symmetry ($s$-symmetric with opposite signs in the two layers). However, in all of the results we present below, the ground state was found to obey $\Delta_{1,\mathbf{k}} = \Delta_{2,\mathbf{k}}$. 

Solving the gap equations self-consistently at a given $\delta$ yields the pairing symmetry of the superconducting ground state. In Fig.~\ref{fig:phase_diagram}, we show the phase diagram as a function of hole doping and of the ratio of real interactions, $J/t$. While $J/t$ is not well characterized for nickelate systems, we assume from the cuprate analogy we find throughout our study that it will not differ strongly from the familiar cuprate values, and hence we consider a relatively wide range $0.15 \le J/t \le 0.35$ centered on the canonical cuprate value of 1/4. For the hole doping, we subtract out the very large value of $\delta_\alpha$, on the assumption that this extremely overdoped band plays little or no role in the superconductivity, and show our results as a function of $\delta_\beta$.

\begin{figure}[t]
\centering
\includegraphics[width=0.85\linewidth]{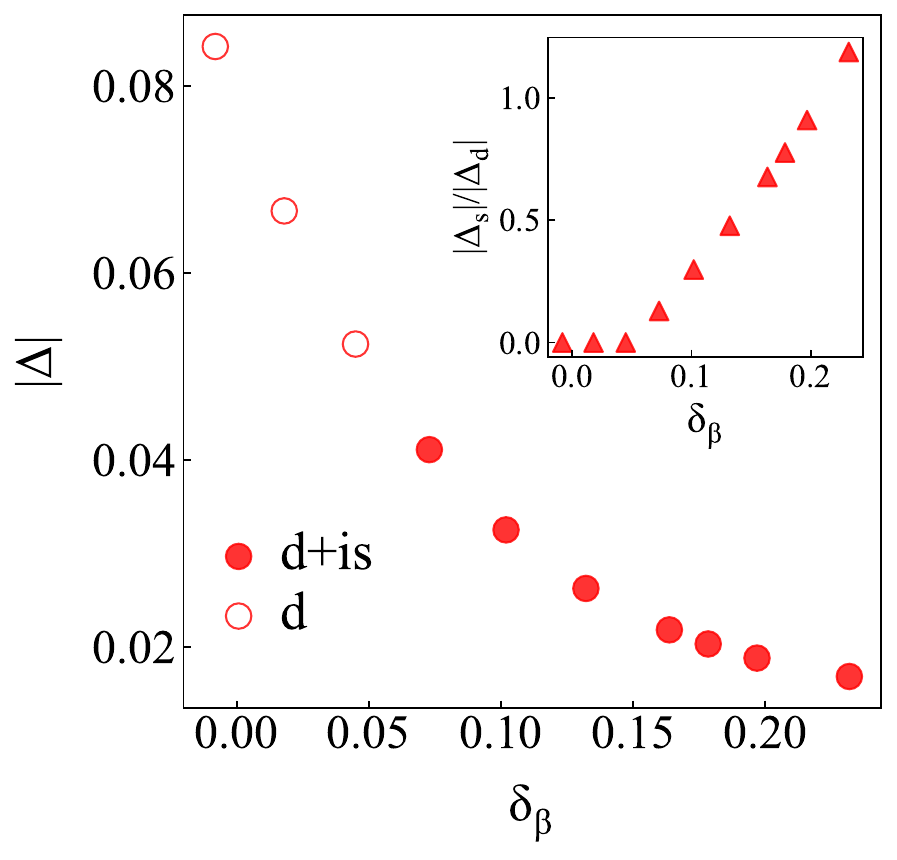}
\caption{Magnitude of the superconducting energy gap, $|\Delta|/t$, shown as a function of $\delta_\beta$ for the renormalized mean-field Hamiltonian with $J / t = 0.3$. Solid circles represent a ($d$+$is$)-wave ground state, and open circles a pure $d$-wave one. The inset shows the ratio between the gap amplitudes of the $s$- and $d$-wave components, $\Delta_s$ and $\Delta_d$, of the ($d$+$is$)-wave state as a function of $\delta_\beta$.} 
\label{fig:phase_cut}
\end{figure}

In this broad parameter range, we find that the superconducting ground state may exhibit either $d$- or ($d$+$is$)-wave pairing symmetry (Fig.~\ref{fig:phase_diagram}). At low net hole concentrations, down to half-filling of the $\beta$ band, pure $d$-wave pairing is favored at all but the highest values of $J/t$. However, increasing the hole content favors an additional $s$-wave component, such that a ($d$+$is$)-wave state appears over much of the phase diagram. Figure~\ref{fig:phase_cut} shows how the gap amplitude varies with $\delta_\beta$ for $J/t = 0.3$, where the $s$-wave component sets in above $\delta_\beta = 0.05$ and, as the inset makes clear, grows systematically with the doping to become comparable with the $d$-wave component at high $\delta_\beta$. Regarding the overall gap amplitude, $|\Delta|$ is very robust at lower dopings and higher $J/t$, approaching the cuprate values of 20-40 meV. 

All of these observations are consistent with the scenario that the near-optimally doped $\beta$ band strongly favors $d$-wave pairing symmetry, as in the highest-$T_c$ cuprates, while the effect of the $\alpha$ band in favoring an additional $s$-wave component becomes increasingly important as the net doping is increased. This ($d$+$is$)-wave pairing state, if substantiated by calculations more rigorous than our present ones, then stands as a key feature distinguishing nickelate superconductivity from the $d$-wave cuprate superconductivity realized in the highest-$T_c$ materials. The pairing symmetry realized in the cuprates \ce{La2CaCu2O6}, \ce{Pb2Sr2YCu3O8}, and \ce{EuSr2Cu2NbO8} may also deviate from pure $d$-wave because both the $\alpha$ and $\beta$ band-fillings deviate quite significantly from optimal. The ($d$+$is$)-wave state has two characteristic properties that can be measured in experiment \cite{RevModPhys.63.239, PhysRevLett.102.217002, PhysRevB.102.220501}. One is the fact that the ($d$+$is$)-wave gap has no node, in contrast to a $d$-wave gap, such that the ``V-shaped'' single-particle tunneling spectrum of a $d$-wave superconductor \cite{Gu2020} is replaced by a ``U-shaped'' spectrum for ($d$+$is$) symmetry. The second is that the ($d$+$is$)-wave state breaks time-reversal symmetry, which can be detected by $\mu$SR measurements in zero field \cite{PhysRevLett.115.267001}.

Considering the doping in more detail, the Ni 3$d_{x^2-y^2}$ and O 2$p_{x,y}$ character of the $\alpha$ and $\beta$ bands means that $\delta$ is to a good approximation the number of 3$d_{x^2-y^2}$ holes (Zhang-Rice singlets) in a single NiO$_2$ plane. The $\gamma$ band consists predominantly of Ni 3$d_{z^2}$ and apical O 2$p_z$ orbitals, and thus has no role in this calculation when it does not cut the Fermi surface. However, the fact that the $\gamma$ band emerges above the Fermi level under pressure, forming a hole pocket around the M point, is important because it extracts holes from the $\alpha$ and $\beta$ bands, allowing the value of $\delta$ to be altered by the applied pressure. The Fermi-surface volume of the $\beta$ band of \ce{La3Ni2O7} at ambient pressure yields an approximate hole doping $\delta_\beta = 0.26$, which corresponds to the heavily overdoped regime in cuprates, where superconductivity is almost completely suppressed. The emergence of the $\gamma$ band above the Fermi level at the pressure represented in Fig.~\ref{fig:cula}(a) reduces $\delta_\beta$ to 0.15, bringing the $\beta$ band into the optimal doping regime that we conclude drives the high-$T_c$ pairing in this material. 

This straightforward reasoning also accounts for the absence of superconductivity in \ce{La3Ni2O6}, which has $\alpha$ and $\beta$ Fermi surfaces comparable to those of \ce{La3Ni2O7} at ambient pressure \cite{zhang2023electronic, PhysRevB.89.224505}, meaning with the $\beta$ band well in the overdoped regime. We draw attention here to the implications of possible O nonstoichiometry in \ce{La3Ni2O7}, in that the presence of O vacancies could cause a rapid suppression of $T_c$ and thus be a source of sample inhomogeneity. Similar to \ce{La3Ni2O6}, the cuprate superconductors with comparable Fermi surfaces [Figs.~\ref{fig:cula}(b-d)] do not possess a $\gamma$ band, and thus neither of the $\alpha$ or $\beta$ bands is close to optimal doping, presumably accounting for the lower $T_c$ values of superconductivity in these systems. 

\section{Summary}
\label{sum}

We have deduced a bilayer $t$-$t_\perp$-$J$ framework to explain the high-temperature superconductivity found in bilayer \ce{La3Ni2O7} at high pressure. Our model is based on two key observations. The first is that the $\alpha$ and $\beta$ bands that dominate the Fermi-surface topology are composed predominantly of Ni 3$d_{x^2-y^2}$ and in-plane O 2$p_x$ and 2$p_y$ orbitals at low energies. The second is that the exact 180$^\circ$ structure of the bilayer leads to an effective quenching of the Hund coupling between the Ni 3$d_{x^2-y^2}$ and 3$d_{z^2}$ orbitals, which arises from the strong interactions within their spatially orthogonal subsystems (respectively the planar NiO$_2$ 3$d_{x^2-y^2}$-2$p_{x,y}$ network and the $c$-axis $3d_{z^2}$-2$p_z$-3$d_{z^2}$ units), and hence to a near-complete decoupling. An essential consequence of the 3$d_{x^2-y^2}$ symmetry of the single active low-energy orbital is the $\mathbf{k}$-dependence of the interlayer tunneling, and the resulting degeneracy of the $\alpha$ and $\beta$ bands along the nodal directions in the Brillouin zone is an important fingerprint of this physics. Other orbitals (such as Ni 4$s$) and processes in the system mean that the prefactor of the interlayer hopping term is nevertheless substantial, which is the origin of the qualitative change in Fermi-surface topology from the best-known bilayer cuprates. However, we observe that the Fermi surfaces of the $\alpha$ and $\beta$ bands in the high-pressure phase of \ce{La3Ni2O7} are very similar to several cuprate materials with strong bilayer coupling, specifically \ce{La2CaCu2O6}, \ce{Pb2Sr2YCu3O8}, and \ce{EuSr2Cu2NbO8}, where the $t$-$J$ model is expected to describe the high-temperature superconductivity.

By solving the superconducting gap equations of the bilayer $t$-$t_\perp$-$J$ model within the renormalized mean-field approximation over a wide range of electron fillings and interaction ratios, we find that the pairing order parameter may have either $d$- or ($d$+$is$)-wave symmetry. We assume that the interaction ratio, $J/t$, is similar to that in cuprates, and find an increase of the gap amplitude with stronger $J$. By decomposing the hole doping into the contributions from the $\alpha$ and $\beta$ bands, we deduce that the high-$T_c$ superconductivity and $d$-wave pairing are driven by the near-optimal doping of the $\beta$ band. An essential contribution to optimizing this doping is made by the $\gamma$ band, which removes holes into the Ni 3$d_{z^2}$ orbitals and provides the mechanism by which the pressure controls the doping level. The pressure (doping) then induces a transition from $d$-wave to ($d$+$is$)-wave symmetry, which at the mean-field level of our calculations appears to be favored as the pairing symmetry in \ce{La3Ni2O7}. This ($d$+$is$)-wave state then stands as the leading difference between nickelate superconductivity and the purely $d$-wave superconductivity realized in the highest-$T_c$ cuprates (\ce{YBa2Cu3O7}$_{-\delta}$, \ce{Ba2Sr2CaCu2O8}, \ce{HgBa2CaCu2O6}, and \ce{Tl2Ba2CaCu2O8}). The ($d$+$is$)-wave state is nodeless and breaks time-reversal symmetry, allowing its verification by a number of experimental methods that can be applied under pressure. 

Concluding with the broader experimental context, bilayer nickelate superconductivity remains a challenging field of research due to the high pressures required to drive the structural transition to the $Fmmm$ phase. Early results showed a very low superconducting volume fraction, which appears to have been a consequence of sample inhomogeneity, and recent experiments suggest that ensuring a robust component of the bilayer (2222) phase with good O stoichiometry does solve the problem of superconducting fragility encountered in first-generation experiments. We reiterate that our minimal $t$-$t_\perp$-$J$ model is formulated for the bilayer phase, which we believe to be the only one allowing 80 K superconductivity; this occurs for the same reason as in cuprates, and the strong bilayer coupling plays a vital role in finding the same physics in a nickelate. The new generation of samples with high structural quality, the increasing development of high-pressure experimental techniques, and further efforts to mimic high physical pressure by the use of substrates or chemical pressure offer a promising near-term future for nickelate superconductivity.

\section*{Acknowledgments}
This work was supported by the National Natural Science Foundation of China (under Grant No.~11888101) and the Postdoctoral Science Foundation of China (under Grant No.~2022M723355).

\begin{appendix}

\section{DFT Calculations} 
\label{ap:dft}

\begin{figure}[t]
\centering
\includegraphics[width=0.75\linewidth]{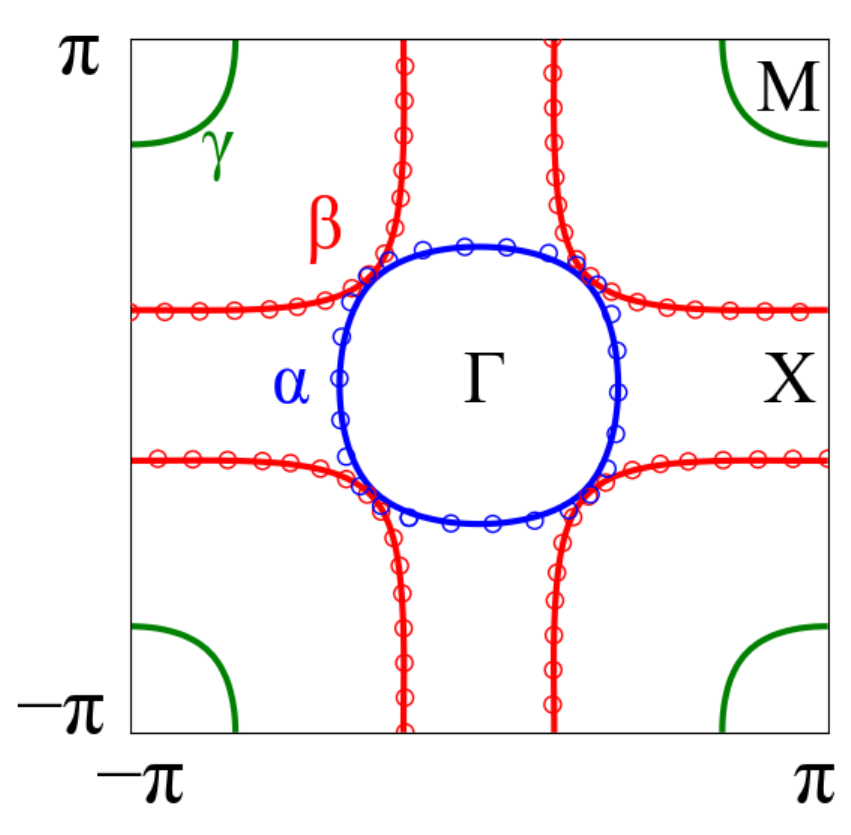}
\caption{Comparison of the Fermi surface obtained from the tight-binding Hamiltonian ($H_0$) with that calculated for \ce{La3Ni2O7} at an effective pressure of 30 GPa by DFT. The solid lines show the DFT Fermi surface, and the open circles show Fermi-surface points obtained from $H_0$ with the parameters given in Sec.~\ref{SubSec:t-J}.}
\label{fig:fit}
\end{figure}

In order to calculate the band structure of \ce{La3Ni2O7} under pressure, and to compare it with the cuprates La$_2$CaCu$_2$O$_6$, Pb$_2$Sr$_2$YCu$_3$O$_8$, and EuSr$_2$Cu$_2$NbO$_8$ as shown in Fig.~\ref{fig:cula}, we performed first-principles DFT calculations for all four materials. For our calculations, we employed the projector augmented wave (PAW) method \cite{paw} implemented in the VASP package~\cite{vasp1,vasp2}. A generalized gradient approximation (GGA) of the Perdew-Burke-Ernzerhof (PBE) type \cite{pbe} was chosen for the exchange-correlation functional. The kinetic energy cutoff of the plane-wave basis was set to 500 eV, and a 12$\times$12$\times$12 $\mathbf{k}$-space mesh was used for sampling the Brillouin zone. A Gaussian broadening with a width of 0.05 eV was implemented for the determination of the Fermi surface (illustrated in Fig.~\ref{fig:fit}). We used the GGA+U formalism of Ref.~\cite{ldau} to incorporate the effect of electron correlations, setting the effective Hubbard $U$ to 3 eV for Ni in \ce{La3Ni2O7} (similar to the value of 3.5 eV used in Ref.~\cite{yang2023orbitaldependent}) and to 6 eV for all three of the cuprates \cite{CuprateU, CuprateU2}. 

Our calculations yielded the electronic band structures of the type shown in Fig.~\ref{fig:cula}(e) for \ce{La3Ni2O7} at a pressure of approximately 30 GPa and for each of the three cuprates at ambient pressure. The lattice constants were fully optimized in every case, with the internal atomic positions relaxed until the forces on all atoms were smaller than 0.01 eV/\AA. To generate the band dispersions of the type depicted in Fig.~\ref{fig:cula}(f), we diagonalized the Hamiltonian, $H_0$ [Eq.~(\ref{eq:2tb})], with parameters determined by fitting the DFT band structures of the $\alpha$ and $\beta$ bands in the energy interval [$-0.4$ eV, 0.4 eV] around the Fermi level. The values of these fitting parameters are given in the main text for \ce{La3Ni2O7} under pressure. Figure \ref{fig:fit} illustrates how this procedure reproduces accurate Fermi-surface contours for the DFT calculations. We note that our results accurately reproduce the giant bilayer splitting in all four materials, consistent with recent results for \ce{La3Ni2O7} \cite{PhysRevLett.131.126001, gu2023effective, zhang2023electronic, sakakibara2023possible, lechermann2023electronic, yang2023minimal} and earlier results for the three bilayer cuprates \cite{PhysRevB.89.224505}.

\end{appendix}

\bibliography{ref}
\end{document}